\documentstyle[preprint,aps,floats,epsfig]{revtex}
\draft
\tightenlines
\begin{document}
\title{
A Multi-Phase Transport model for nuclear collisions at RHIC
}
\bigskip
\author{
Bin Zhang\footnote{Email: bzhang@kopc1.tamu.edu}$^{\rm a}$,
C.M. Ko\footnote{Email: ko@comp.tamu.edu}$^{\rm a}$,
Bao-An Li\footnote{Email: bali@navajo.astate.edu}$^{\rm b}$,
and Ziwei Lin\footnote{Email: lin@kopc1.tamu.edu}$^{\rm a}$}
\address{$^{\rm a}$Cyclotron Institute and Physics Department,\\
Texas A\&M University, College Station, TX 77843, USA}
\address{$^{\rm b}$Department of Chemistry and Physics\\
Arkansas State University, P.O. Box 419\\
State University, AR72467-0419, USA}
%\date{June 30, 1999}
\maketitle

\begin{abstract}
To study heavy ion collisions at energies available from the
Relativistic Heavy Ion Collider, we have developed a
multi-phase transport model that includes both initial 
partonic and final hadronic interactions. Specifically, the parton 
cascade model ZPC, which uses as input the parton distribution from 
the HIJING model, is extended to include the quark-gluon to hadronic matter
transition and also final-state hadronic interactions based on the
ART model. Predictions of the model for central Au on Au collisions 
at RHIC are reported.
\end{abstract}
%\noindent{PACS number(s): 25.75.Ld, 13.75.Jz, 21.65.+f}
\pacs{25.75.-q, 24.10.Lx, 24.10.Jv}

% Introduction
The beginning of experiments at the Relativistic Heavy Ion Collider 
(RHIC) this year will start an exciting new era in nuclear
and particle physics. The estimated high energy density in central
heavy ion collisions at RHIC is expected to lead to the formation of a large
region of deconfined matter of quarks and gluons, the Quark Gluon
Plasma (QGP). This gives us an opportunity to study the properties
of QGP and its transition to the hadronic matter, which would then
shed light on the underlying fundamental theory of strong
interactions, the Quantum Chromodynamics (QCD).  

Because of the complexity of heavy ion collision dynamics, Monte Carlo event 
generators are needed to relate the experimental observations to the 
underlying theory. This has already been shown to be the case in heavy 
ion collisions at existing accelerators such as the SIS, AGS, and SPS 
\cite{koli96,art1,kahana96,cass99,sorge95,bass99}. 
As minijet production is expected to play an important role at RHIC
energies \cite{hijing1}, models for partonic transport have already
been developed \cite{geiger,zpc1}. Furthermore, transport models that 
include both partonic and hadronic degrees of freedom are being developed 
\cite{ynara1,sbass1}. We have recently also developed such a
multi-phase transport (AMPT) model.  It starts from initial
conditions that are motivated by the perturbative QCD and
incorporates the subsequent partonic and hadronic space-time
evolution. In particular, we have used the HIJING model
\cite{hijing1} to generate the initial phase space
distribution of partons and the ZPC model \cite{zpc1} to follow
their rescatterings. A modified HIJING fragmentation scheme is then
introduced for treating the hadronization of the partonic matter. 
Evolution of the resulting hadron system is treated in the framework
of the ART transport model \cite{art1}. In this paper, we shall 
describe this new multi-phase transport model and show its
predictions for central Au on Au collisions at RHIC.

% Elements of AMPT
In the AMPT model, the initial parton momentum distribution is 
generated from the HIJING model, which is a Monte-Carlo event generator 
for hadron-hadron, hadron-nucleus, and nucleus-nucleus collisions. 
The HIJING model treats a nucleus-nucleus collision as a superposition
of many binary nucleon-nucleon collisions. For each pair of nucleons,
the impact parameter is determined using nucleon transverse
positions generated from a Wood-Saxon nuclear density distribution.
The eikonal formalism is then used to determine the probability for a
collision to occur. For a given collision, one further determines
if it is an elastic or inelastic collision, a soft or hard
inelastic interaction, and the number of jets produced in a hard
interaction. To take into account nuclear effects in hard
interactions, an impact parameter-dependent parton distribution function 
based on the Mueller-Qiu parameterization \cite{amueller1}
of the nuclear shadowing is used. Afterwards, PYTHIA routines 
\cite{sjostrand1} are called to describe hard interactions, while 
soft interactions are treated according to the Lund model 
\cite{bandersson1}.

In HIJING model, minijets from produced partons are quenched by 
losing energy to the wounded nucleons close to their straight-line 
trajectories. In the AMPT model, we replace the parton quenching
by their rescatterings. To generate the initial phase space distribution 
for the parton cascade, the formation time for each parton is determined 
according to a Lorentzian distribution with a half width $t_f=E/m_T^2$ 
\cite{gyulassy1}, where $E$ and $m_T$ are the parton energy and transverse 
mass, respectively.  Positions of formed partons are calculated from 
those of their parent nucleons using straight-line trajectories.
Since partons are considered to be part of the coherent cloud of 
parent nucleons during the time of formation, they thus do not suffer 
rescatterings. 

The parton cascade in the AMPT model is carried out using the ZPC
model \cite{zpc1}. At present, this model includes only gluon-gluon elastic
scatterings with cross sections taken to be the leading
divergent cross section regulated by a medium generated screening
mass. The latter is related to the phase space density of produced
partons \cite{tbiro1}. In the present study, a constant screening mass
of $\mu=3\, {\rm fm}^{-1}$ is used.

Once partons stop interacting, they are converted into hadrons
using the HIJING fragmentation scheme after an additional proper
time of approximate $1.2$ fm. In the default HIJING model, a diquark 
is treated as a single entity, and this leads to an average rapidity 
shift of about one unit in the net baryon distribution. We modify this
fragmentation scheme to allow the formation of diquark-antidiquark
pairs. In addition, the $BM\bar{B}$ formation probability is 
taken to be $80\%$ for the produced diquark-antidiquark pairs, while 
the rest are $B\bar{B}$'s. This gives a reasonable description of the
measured net baryon rapidity distribution in Pb+Pb collisions at
158 GeV/nucleon \cite{bzhang1}.

For the evolution of hadrons, we use the ART model, which is a 
successful hadronic transport model for heavy ion collisions at AGS
energies. To extend the model to heavy ion collisions at RHIC, we
have further included nucleon-antinucleon annihilation channels, the 
inelastic interactions of kaons and antikaons, and neutral kaon production.
In the ART model, multiparticle production is modeled through the
formation of resonances. Since the inverse double resonance
channels have smaller cross sections than those calculated directly
from the detailed balance, we thus adjust the double resonance
absorption cross sections to fit the NA49 data \cite{bzhang1}.

% Results and discussions

\begin{figure}[ht]
\centerline{\epsfig{file=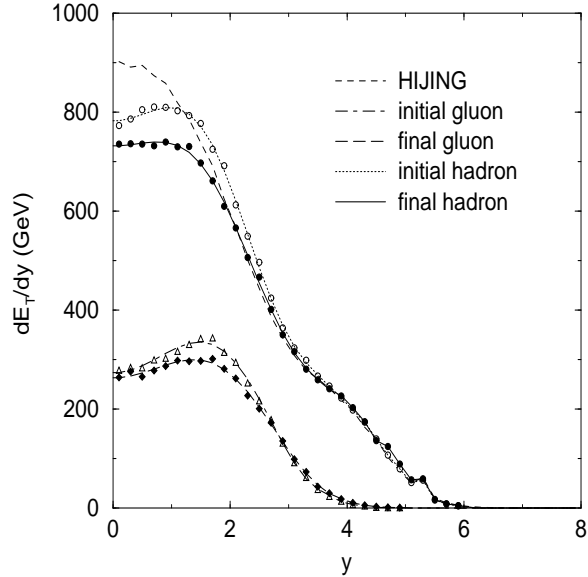,height=3.0in,width=3.0in,angle=-90}}
\vspace{0.3in}
\caption{
Transverse energy rapidity distribution for
central (b=0) Au+Au collisions at RHIC.
\label{fig:rhic1}}
\end{figure}

In Fig. \ref{fig:rhic1}, we show the rapidity distribution of 
transverse energy in Au+Au central ($b=0$) collisions at RHIC.
In the default HIJING model, the $dE_T/dy$ at central rapidity 
is about 900 GeV. Using the modified HIJING framentation scheme leads 
to a decrease of about 100 GeV in $dE_T/dy$. After the
parton cascade, $dE_T/dy$ is reduced by about 15 GeV as shown
by the difference in the initial and final gluon $dE_T/dy$ 
distributions. We note that the perturbatively produced gluons account for a 
significant fraction (about 1/3) of the produced $dE_T/dy$.
Including hadronic evolution further reduces $dE_T/dy$ by 
about 50 GeV.  As the transverse energy rapidity distribution is a 
sensitive probe of longitudinal expansion \cite{gyulassy2} and 
is related to the $pdV$ work, these results indicate that both the 
partonic and hadronic evolution contribute appreciably
to longitudinal collective flow.

\begin{figure}[ht]
\centerline{\epsfig{file=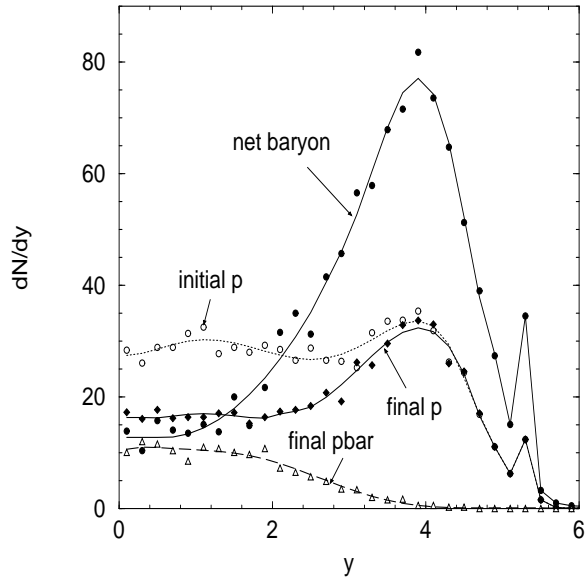,height=3.0in,width=3.0in,angle=-90}}
\vspace{0.3in}
\caption{Baryon rapidity distributions for
RHIC Au+Au central (b=0) collisions.
\label{fig:rhic2}}
\end{figure}

Fig. \ref{fig:rhic2} shows the baryon rapidity distributions.
It is seen that the net baryon distribution from the AMPT model has a 
peak value of 80 at $y\sim 3.9$ while that from the default HIJING 
model has a peak value of 85 at $y\sim 4.5$. The larger rapidity shift 
in AMPT is due to the modified fragmentation of diquarks. At 
central rapidity, AMPT predicts a net baryon number of about 12,
which is similar to that from the default HIJING model.
Many antiprotons (about $50\%$) are seen to survive the
absorption in hadronic matter, leading to a value of about 10
at central rapidities. The $\bar{p}/p$ ratio at central rapidity is
about $60\%$, which is much larger than the $10\%$ seen in 
Pb+Pb collisions at 158 GeV/nucleon from SPS \cite{nxu1}.

\begin{figure}[ht]
\centerline{\epsfig{file=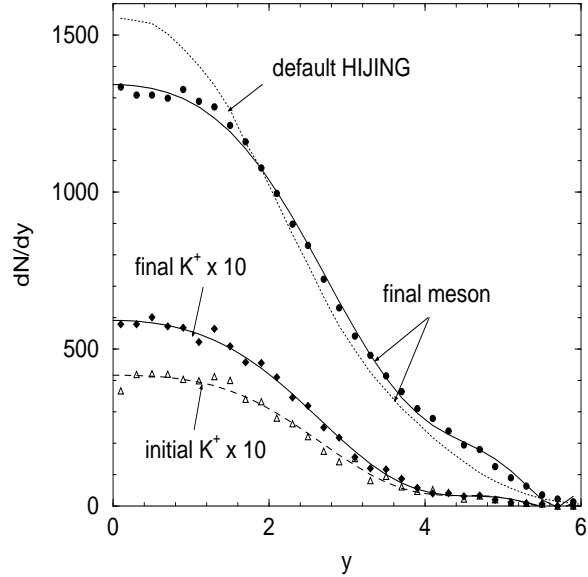,height=3.0in,width=3.0in,angle=-90}}
\vspace{0.3in}
\caption{Meson rapidity distributions for
RHIC Au+Au central (b=0) collisions.
\label{fig:rhic3}}
\end{figure}

The final meson rapidity distribution is shown in Fig.\ref{fig:rhic3}. 
The prediction from the AMPT model has a distinctive plateau structure 
around central rapidities.  Results using the default HIJING model show
instead a peak at central rapidity with a higher rapidity density.
Also shown in the figure is the distribution of kaons produced from both 
string fragmentation and hadronic interaction. The latter is seen 
to enhance significantly the kaon yield.

\begin{figure}[ht]
\centerline{\epsfig{file=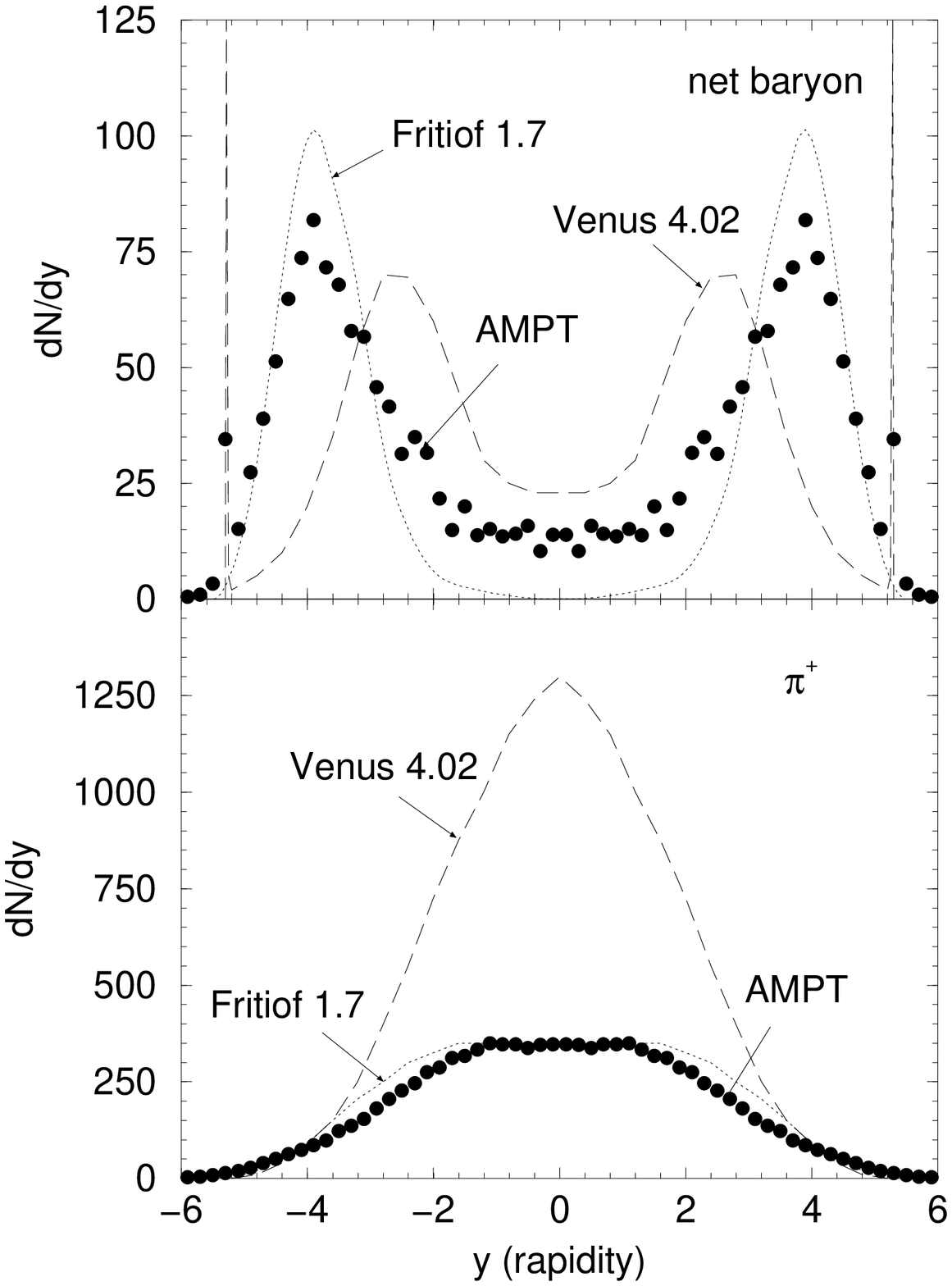,height=4.2in,width=3.2in,angle=0}}
\vspace{0.3in}
\caption{Comparisons of AMPT, Fritiof1.7, and Venus4.02 predictions
for RHIC Au+Au central (b=0) collisions.
\label{fig:rhic4}}
\end{figure}

In Fig. \ref{fig:rhic4}, the AMPT results are compared with those from 
other models without final-state rescatterings such as the Fritiof1.7 
and the Venus4.02 \cite{dbeavis1}.  Although both AMPT and Fritiof show
similar peaks at $y\sim 3.9$ in the net baryon rapidity distributions,
the height of the peak is about 80 in AMPT but is about 100 in Fritiof.  
On the other hand, the net baryon distribution peaks at a smaller
rapidity of $y\sim 2.8$ in Venus.  At central rapidities,
the net baryon number from AMPT is about 15 and is similar to that
from Venus but is much larger than that from Fritiof, which 
is almost zero at the central rapidity. For mesons, both AMPT and 
Fritiof have final $\pi^+$ rapidity distributions that peak at the 
central rapidity with a height of about 350 while Venus gives a much 
larger height of about 1300 at the central rapidity.
The forthcoming RHIC data will allow us to test the different predictions
from these models and thus to obtain a better understanding of the 
collision dynamics. 

% Conclusions

In conclusion, we have developed for heavy ion collisions at 
the Relativistic Heavy Ion Collider a multi-phase transport model 
that includes both partonic and hadronic evolution.  The model
shows that both partons and hadrons contribute to the longitudinal 
collective work. Because of the 
production of diquark-antidiquark pairs, there is a relatively large 
rapidity shift of net baryons compared to the default HIJING fragmentation
scheme. Many anti-protons survive final-state interactions and
are expected to be observed at RHIC. Also, our model gives a wider meson
rapidity plateau at central rapidities than the prediction from
the default HIJING model. Furthermore, kaon production is appreciably
enhanced due to production from hadronic interactions.
For future studies, we shall compare these predictions with the 
experimental data soon to be available from RHIC. Also, we shall 
study if inclusion of parton inelastic scatterings and 
using different hadronization schemes would affect the results
obtained here.

This work is supported by the National Science Foundation under Grant 
No. PHY-9870038, the Welch
Foundation Grant A-1358, and the Texas Advanced Research Program
FY97-010366-068.

%%%%%%%%%%%%%%%%%%%%%%%%%%%%%%%%%%%%%%%%%%%%%%%%%%%%%%%%%%%%%%%%%

\end{document}